\documentclass[twocolumn,pra,amsmath,amssymb]{revtex4-2}
\usepackage{amsmath,amssymb,graphicx,bm,epsfig,color}
\usepackage{qcircuit}
\usepackage{physics}
\usepackage{subfig}
\begin{document}
\title{ Bell inequality violation on small NISQ computers}

\author{H.W.L. Naus}
\author{H. Polinder} \email{henk.polinder@tno.nl}
\affiliation{Quantum Technology, Netherlands Organization for Applied Scientific Research (TNO), Delft, The Netherlands}

\begin{abstract}
Quantum computational experiments exploiting Noisy Intermediate-Scale Quantum (NISQ) devices to demonstrate violation of a Bell inequality are proposed.
They consist of running specified quantum algorithms on few-qubit computers.
If such a device assures entanglement and performs single-shot measurements, the detection
loophole is avoided. 
Four concise quantum circuits determining the expectation values of
the relevant observables are used for a two-qubit system. 
It is possible to add an ancilla qubit to these circuits and eventually only measure the ancilla
to obtain the relevant information.
For a four-qubit NISQ computer, two algorithms yielding the same averages, however
also guaranteeing a random choice of the observable, are developed.
A freedom-of-choice loophole is therefore avoided.
Including an additional ancilla 
reduces the number of measurements by one since in this case only the ancillas need to be measured.
Note that these methods, using the NISQ device, are intrinsically quantum mechanical.
Locality loopholes cannot be excluded
on present NISQ systems. Results of simulations on the QX simulator of Quantum Inspire 
are presented. The Bell inequality is indeed found to be violated, even if some
additional noise is included by means of the depolarizing channel error model.
The algorithms have been implemented on the IBM Q Experience as well. The results of 
these quantum computations support a violation of the Bell inequality by various standard
deviations.   
\end{abstract}

\date{\today}
\maketitle

\newcommand{\tf}{\tfrac{1}{2}\sqrt{2}}
\newcommand{\ts}{\tfrac{1}{2}}


\section{Introduction}

Noisy Intermediate-Scale Quantum (NISQ) computers are becoming available 
and the question arises what kind of useful computations can be done with such devices \cite{Preskill}.
One aims to exploit typical quantum properties such as superposition and entanglement.
The simplest system where entanglement is possible of course has two qubits. 
Quantum mechanical entanglement is essential for the famous Einstein-Podolsky-Rosen (EPR) paradox
\cite{EPR}. Bell \cite{Bell} addressed this paradox by formulating inequalities, now known as
Bell inequalities \cite{Yurke,Zuko}. As pointed out in \cite{NC}, they are based on `common sense' notions how
nature is supposed to behave. Quantum mechanics, however, appears to violate these
inequalities, see e.g. \cite{Hensen,Vedo,Rauch} and references therein.

A simple form of a Bell test has originally been proposed in \cite{CHSH}. 
Here we follow its presentation as given in \cite{NC}. It indeed exploits a
two qubit system, implying that the quantum mechanical experiment can in
principle be done. In contrast to \cite{Hensen,Bernien}, spatial separation will not be feasible
with present NISQ computers - it may be possible in the future.
In contrast to the locality loophole, the detection loophole is avoided by
a quantum computer guaranteeing entanglement and enabling single-shot
measurements.  


In this paper, we first show how to implement the proposed quantum mechanical experiment
on a two-qubit quantum computer. It is assumed that only measurement of the $Z$ Pauli operator can
be made; it is called measuring in the standard basis. Of course, a measurement in the
$X$-basis can be done by using an additional Hadamard operation and a standard measurement.
Here we also derive the operations necessary for determining the expectation values
of the other two involved observables $\tf(\pm Z-X) $. Hence we present the four quantum circuits for
this Bell experiment.
Next, the experiments are extended with a quantum mechanically random choice
of the observable to be measured in each run. In this way, a free and random
measurement setting is obtained, thereby  evading the ``freedom-of-choice'' loophole  \cite{Rauch,Weihs,Scheidl,Piro}.
To this end, two additional qubits, `ancillas', 
are added, requiring a four-qubit device. Its implementation on IBM Q Experience \cite{IBM},
quantum computing with superconducting qubits, can therefore be seen as an extension of \cite{Ansmann}.  
Next, the experiments for a two- and the four-qubit system are extended with an additional ancilla. By measuring the
ancilla(s) {\it only}, the relevant information is obtained.

In the Quantum Inspire project \cite{QI} the simulator QX is available.  
Hence we can simulate the proposed quantum experiments by implementing
the circuits mentioned above. Running these quantum programs a number of times approximates the desired expectation values.
Our algorithms are 
implemented and tested on IBM Q Experience as well. In this way, the violation of
Bell's inequality is addressed on a real NISQ computer, avoiding the detection and
freedom-of-choice loopholes. We cannot avoid the locality loophole in this way.
The experiments can also be seen as severe test of such a quantum device, in particular
with respect to entanglement, decoherence and single-shot measurements.

\section{Bell inequality and its violation in quantum mechanics}
To make this paper self-contained we briefly recapitulate some results for the experiment
proposed in \cite{CHSH}, although essentially following \cite{NC}, which contains more details than we present here.

The derivation of the Bell inequality for the classical experiment uses four properties  of particles,
$Q, R, S, T$ which are supposed to be objective. At one location either $Q$ or $R$ is measured, at 
another location $S$ or $T$. These measurements are to be performed simultaneously.
The outcomes are always $+1$ or $-1$. Then, we see from 
\begin{equation}
QS+RS+RT-QT = (Q+R)S+(R-Q)T
\end{equation} 
that either $(Q+R)S = 0$ or $(R-Q)T = 0$. 
It follows that $QS+RS+RT-QT = \pm 2$.
The probability that the system is in a state with
$Q=q, R=r, S=s,T =t $ is given by $p(q,r,s,t)$. 
Note that a joint probability distribution is used, {\it i.e.}, independence of the variables
is {\it not} assumed. For the expectation $E[\cdot]$ two relations
can be derived:
\begin{align}
&E[QS+RS+RT-QT] = \nonumber \\ &\sum_{qrst} p(q,r,s,t) (qs+rs+rt-qt) \nonumber \\
&\le \sum_{qrst} p(q,r,s,t) \times 2 = 2
\end{align}
and
\begin{align}
&E[QS+RS+RT-QT] = \nonumber \\ & \sum_{qrst} p(q,r,s,t) qs
+ \sum_{qrst} p(q,r,s,t) rs \nonumber \\
&+ \sum_{qrst} p(q,r,s,t) rt
- \sum_{qrst} p(q,r,s,t) qt \nonumber \\
&= E[QS] + E[RS]+ E[RT] - E[QT].
\end{align}
Combining these relations immediately yields the famous Bell inequality
\begin{equation}
E[QS] + E[RS]+ E[RT] - E[QT] \le 2.
\label{eq:Bellin}
\end{equation} 
It is also known as the CHSH inequality \cite{CHSH}. 

The concomitant quantum experiment using two qubits starts with the 
preparation of the Bell state
\begin{equation}
| \Psi \rangle  = \tf (\ket{01} - \ket{10}).
\label{eq:Bell}
\end{equation} 
The following observables are considered
\begin{eqnarray}
Q &=& Z_1, \; R = X_1, \; S=-\tf \left(Z_2+X_2\right), \nonumber \\
 T &=& \tf \left(Z_2 - X_2\right).
\label{eq:operators}
\end{eqnarray} 
In order to produce their quantum mechanical expectation values,
the products $QS, RS, RT, QT$ are repeatedly measured.
With standard  quantum mechanical calculations, one readily obtains the results
for this Bell state:
\begin{eqnarray}
< QS >  &=& <RS>  = <RT> = \tf \quad , \nonumber \\
 <QT> &=& - \tf
\label{eq:evalues}
\end{eqnarray} 
and, consequently
\begin{equation}
< QS >  + <RS>  + <RT> -  <QT> = 2\sqrt{2}.
\label{eq:evalues1}
\end{equation} 
Comparison of the Bell inequality (\ref{eq:Bellin}) and these theoretical results (\ref{eq:evalues1})
demonstrates the violation of the Bell inequality in quantum mechanics.

We stress that measuring these products of operators in principle
requires {\it simultaneous} measurements of the two qubits.
In practice such measurements take a certain amount of time and
this synchronicity is only approximately obtained.
It is presupposed that approximate simultaneous measurement is possible for the $Z$-operators, {\it i.e.}, in the standard $Z$-basis.
By using an additional ancilla, this requirement is not present since only the ancilla
needs to be eventually measured. 

In the common discussions concerning local realism, it is argued that the distance
between the two qubits should be large enough to exclude mutual influence of the measurements.
The locations should be causally disconnected. 
In the envisaged experiment this is of course not the case since the qubits
are very close to each other, actually on the same chip.
In other words, we cannot avoid the locality loophole. 

Another loophole is the so-called detection loophole, in \cite{Hensen} phrased
as `guaranteeing' efficient measurements. If the quantum device indeed generates
entanglement with certainty and obtains measurements of qubits every time \cite{Ansmann},
no additional `fair sampling' hypothesis is necessary. As a consequence,
our experiments are indifferent to the detection loophole.

The recently \cite{Rauch} re-addressed freedom-of-choice loophole \cite{Weihs,Scheidl, Piro} is avoided by
randomly choosing the operator out of the four introduced observables
to be measured in one run; it is called `random basis choice' in \cite{Hensen}.
We achieve this by exploiting a four qubit-device, implementing the random selection
quantum mechanically by means of two additional qubits. 
Sec. \ref{sec:Bell-III} of this paper presents this novel approach in detail.
  
\section{Bell experiments I}\label{sec:Bell-I}
\subsection{Preparation of the Bell state}\label{PR}
The entangled quantum state (\ref{eq:Bell}) needs to be prepared first from the
standard initial state $\ket{00}$. As is well-known the operator
$\text{CNOT} (H_1\otimes \mathcal I)  (X_2 \otimes X_1)$ does the job;
$H$ denotes the Hadamard gate and CNOT is the entangling controlled-not operation \cite{NC}.
Explicitly we indeed get 
\begin{align}
\text{CNOT} (H_1 \otimes \mathcal I) (X_2 \otimes X_1) \ket{00} &= 
\text{CNOT} (H_1 \otimes \mathcal I)  \ket{11} = \nonumber \\
\tf \text{CNOT} (\ket{0} - \ket{1})\otimes \ket{1} &= \tf(\ket{01}-\ket{10}), 
\end{align}
that is the desired Bell state.

\subsection{Measurements of observables}\label{sec:MQ}
Determining the expectation value $\bra{\Psi} U \ket{\Psi}$
of a Hermitian operator $U$
by eventually measuring $Z$ is done as follows. Here we only consider one-qubit operators, or, equivalently,
two-by-two matrices,
with eigenvalues $\pm 1$. The matrix $U$ has to be diagonalized
by means of a unitary operator $O$:
\begin{equation}
 O U O^\dagger = Z.
\end{equation} 
The diagonalized matrix is the Pauli $Z$ because of its eigenvalues being $\pm 1$.
Then we obtain 
\begin{equation}
\bra{\Psi} U \ket{\Psi} = 
\bra{\Psi} O^\dagger O U O^\dagger O \ket{\Psi} = 
\bra{\Phi} Z \ket{\Phi} ,
\label{eq:dia}
\end{equation} 
where $\ket{ \Phi} = O \ket{\Psi}$.
Linear algebra tells us that the unitary
transformation follows from the eigenvectors of the matrix $U$.
Measuring the Pauli $Z$ in the transformed state $\Phi$ is therefore
statistically  equivalent to measuring 
the observable $U$ in the original state $\ket{\Psi}$.
However, the final `collapsed' states are in general different.

Let us first demonstrate this for the Pauli $X$  operator. The result is of course well-known, the Hadamard
operator $H$ does the job:
\begin{equation}
H X H = Z, 
\end{equation}
as is verified by explicit matrix multiplication. The columns of $H$ are indeed the eigenvectors of $X$.
An $X$ `measurement' requires an additional Hadamard operation followed by measuring $Z$.

The observables used in the proposed quantum Bell experiment also require measurements of the
operators $S$ and $T$ given in  (\ref{eq:operators}). Thus we first have to solve the respective
eigenvalue problems. Omitting these straightforward calculations, we present the resulting unitary 
transformations and verify that they indeed transform $S$ and $T$ to the Pauli $Z$. 
For $S$ we obtain as unitary transformation the rotation
\begin{equation}
R_y(\vartheta)  = \cos{(\ts \vartheta)} \mathcal I- i \sin{(\ts \vartheta)} Y, \; \text{with} \;
\vartheta=  -\frac{5\pi}{4}. 
\end{equation}
Transforming $S$ yields
\begin{eqnarray}
R_y(\vartheta) S R_y(-\vartheta)  &=& \cos^2{(\ts \vartheta)} S - i \sin{(\ts \vartheta)}\cos{(\ts \vartheta)} [Y,S]  
\nonumber \\ &&+\sin^2{(\ts\vartheta)} YSY.   
\end{eqnarray}
Simplifying the commutator 
\begin{equation}
[Y,S] = i \sqrt{2}(Z-X)
\end{equation}
and the triple product as
\begin{equation}
YSY = S + \sqrt{2}(Z+X),
\end{equation}
gives
\begin{align}
&R_y(\vartheta) S R_y(-\vartheta)  = \nonumber \\
&\sqrt{2}\left(\sin^2{(\ts\vartheta)} - \sin{(\ts \vartheta)}\cos{(\ts \vartheta)} -\ts\right) X   \nonumber \\
&+\sqrt{2}(\sin^2{(\ts\vartheta)}+\sin{(\ts \vartheta)}\cos{(\ts \vartheta)}-\ts) Z.    
\end{align}
At this point, we put in the values $\cos{(-\frac{5\pi}{8})}= -\ts \sqrt{2-\sqrt{2}}$
and $\sin{(-\frac{5\pi}{8})}= -\ts \sqrt{2+\sqrt{2}}$ and obtain
\begin{equation}
R_y(\vartheta) S R_y(-\vartheta)  = Z.
\end{equation}

The $T$ observable is diagonalized by the rotation
\begin{equation}
R_y(\alpha)  = \cos{(\ts \alpha)} \mathcal I - i \sin{(\ts \alpha)} Y, \quad \text{with} \quad \alpha= \frac{\pi}{4}. 
\end{equation}
Its verification proceeds analogously 
\begin{eqnarray}
R_y(\alpha) T R_y(-\alpha) &=& \cos^2{(\ts \alpha)} T - i \sin{(\ts \alpha)}\cos{(\ts \alpha)} [Y,T]  \nonumber \\
&&+\sin^2{(\ts\alpha)} YTY. 
\end{eqnarray}
Using
\begin{equation}
[Y,T] = i \sqrt{2}(Z+X)
\end{equation}
and 
\begin{equation}
YTY = T + \sqrt{2}(X-Z),
\end{equation}
yields
\begin{align}
&R_y(\alpha) T R_y(-\alpha)  = \nonumber \\
&\sqrt{2}\left(\sin^2{(\ts\alpha)}+\sin{(\ts \alpha)}\cos{(\ts \alpha)}  -\ts\right) X   \nonumber \\
&+\sqrt{2}\left(\ts - \sin^2{(\ts\alpha)}+\sin{(\ts \alpha)}\cos{(\ts \alpha)}\right) Z.    
\end{align}
By means of
 $\cos{(\frac{\pi}{8})}= \ts \sqrt{2+\sqrt{2}}$
and $\sin{(\frac{\pi}{8})}= \ts \sqrt{2-\sqrt{2}}$ we indeed get
\begin{equation}
R_y(\alpha) T R_y(-\alpha)  = Z.
\end{equation}
This completes the computations necessary for the quantum algorithms corresponding to the Bell experiment.

\subsection{Quantum circuits}\label{QC}
\begin{figure}[ht!]
\centering
	\scalebox{1.65}{
		\begin{tabular}{c}
\Qcircuit @C=1em @R=.8em
{\lstick{\ket{0}}  & \gate{X} & \gate{H} & \ctrl{1} & \qw & \meter \\
\lstick{\ket{0}}  & \gate{X} & \qw & \targ & \gate{R_y(\vartheta)} & \meter} 
		\end{tabular}
}
\caption{QS circuit, $\vartheta=-5\pi/4$.} \label{cir:QS} 
\end{figure}
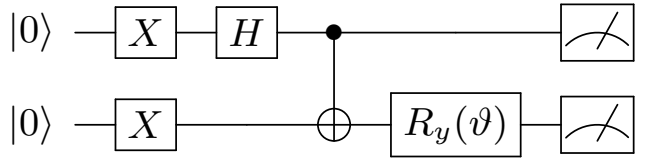

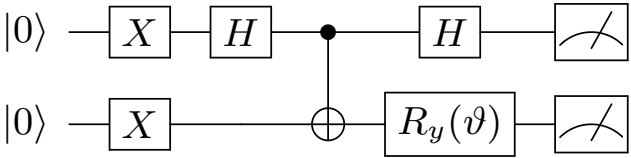
\begin{figure}[ht!]
\centering
	\scalebox{1.65}{
		\begin{tabular}{c}
\Qcircuit @C=1em @R=.8em
{\lstick{\ket{0}}  & \gate{X} & \gate{H} & \ctrl{1} & \gate{H} & \meter \\
\lstick{\ket{0}}  & \gate{X} & \qw & \targ & \gate{R_y(\vartheta)} & \meter} 
		\end{tabular}
}
\caption{RS circuit, $\vartheta=-5\pi/4$.} \label{cir:RS} 
\end{figure}

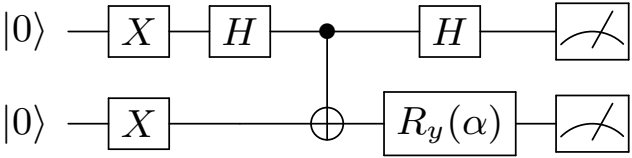
\begin{figure}[ht!]
\centering
	\scalebox{1.65}{
		\begin{tabular}{c}
\Qcircuit @C=1em @R=.8em
{\lstick{\ket{0}}  & \gate{X} & \gate{H} & \ctrl{1} & \gate{H} & \meter \\
\lstick{\ket{0}}  & \gate{X} & \qw & \targ & \gate{R_y(\alpha)} & \meter} 
		\end{tabular}
}
\caption{RT circuit, $\alpha=\pi/4$.} \label{cir:RT} 
\end{figure}

\begin{figure}[ht!]
\centering
	\scalebox{1.65}{
		\begin{tabular}{c}
\Qcircuit @C=1em @R=.8em
{\lstick{\ket{0}}  & \gate{X} & \gate{H} & \ctrl{1} & \qw & \meter \\
\lstick{\ket{0}}  & \gate{X} & \qw & \targ & \gate{R_y(\alpha)} & \meter} 
		\end{tabular}
}
\caption{QT circuit, $\alpha=\pi/4$.} \label{cir:QT} 
\end{figure}
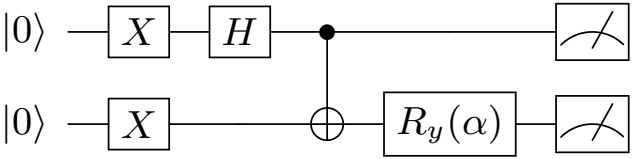

Preparation of the Bell state and the implementation of
the measurements described in Sec. \ref{sec:MQ} yields the quantum circuits,
or, equivalently, quantum algorithms shown in Figs. \ref{cir:QS} -- \ref{cir:QT}.

\section{Bell experiments II: alternative implementation}\label{sec:AI}
\subsection{Single qubit}
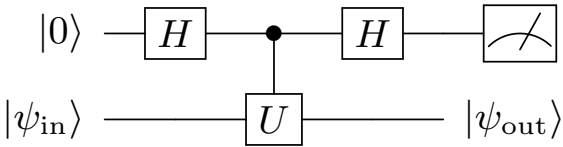
\begin{figure}[ht!]
\centering
	\scalebox{1.65}{
		\begin{tabular}{c}
\Qcircuit @C=1em @R=.8em
{\lstick{\ket{0}}  &  \gate{H} & \ctrl{1} & \gate{H} & \qw & \meter \\
\lstick{\ket{\psi_{\text{in}}}}  &  \qw &  \gate{U} & \qw & \rstick{\ket{\psi_{\text{out}}}} \qw } 
		\end{tabular}
}
\caption{Measuring an operator $U$.} \label{cir:MU} 
\end{figure}

Instead of the implementation by means of the additional rotations described in the previous section, we exploit an additional qubit in order
to measure the appearing operators in this section. The basic circuit for the single qubit in state $\ket{\psi_{\text{in}}}$ is presented in exercise {\bf 4.34} in \cite{NC},
here depicted in Fig. \ref{cir:MU}.

The operator $U$ is hermitian and unitary and therefore has eigenvalues $\pm 1$. Its orthonormal eigenstates $\ket{u_\pm}$
satisfy
\begin{eqnarray}
U \ket{u_\pm} = \pm \ket{u_\pm}, \quad \braket{u_+}&=&\braket{u_-}=1, \nonumber \\ \braket{u_+}{u_-} &=&0.
\end{eqnarray} 
For completeness we prove the assertion of \cite{NC}. Explicitly it needs to be checked
that the measurement probabilities 
\begin{equation}
p_+ =  \left|\braket{u_+}{\psi_{\text{in}}}\right|^2, 
p_- =  \left|\braket{u_-}{\psi_{\text{in}}}\right|^2
\end{equation} 
as well as eigenstates
\begin{eqnarray}
&&U \ket{\psi_{\text{out}}} = \pm \ket{\psi_{\text{out}}}, \quad \text{that is} \nonumber \\
&&\ket{\psi_{\text{out}}} = \ket{u_+} \quad \text{or}
\quad \ket{\psi_{\text{out}}} = \ket{u_-}
\end{eqnarray} 
are reproduced.  
Starting with initial two-qubit state
$\ket{0} \otimes \ket{\psi_{\text{in}}}$, we obtain as state before the measurement
\begin{equation}
\ket{\psi_f} = \tfrac{1}{2} \ket{0} \otimes \left(\ket{\psi_\text{in}} + U \ket{\psi_\text{in}}\right)
+ \tfrac{1}{2} \ket{1} \otimes \left(\ket{\psi_\text{in}} - U \ket{\psi_\text{in}}\right).
\end{equation} 
The measurement of the first qubit, {\it i.e.}, the ancilla, yields the outcomes $0,1$
with respective probabilities
\begin{align}
p_0 &= \tfrac{1}{2}\left(1+\bra{\psi_{\text{in}}} U \ket{\psi_{\text{in}}}\right), \nonumber \\
p_1 &= \tfrac{1}{2}\left(1-\bra{\psi_{\text{in}}} U \ket{\psi_{\text{in}}}\right).
\end{align} 
By means of the spectral representation of the operator $U$,
\begin{equation}
U = \ket{u_+}\bra{u_+} - \ket{u_-}\bra{u_-},
\end{equation} 
and the orthonormality of the states $\ket{u_\pm}$, it indeed follows that $p_0=p_+$ and $p_1=p_-$.
After the measurement with the results $0,1$ the  one-qubit states respectively are
\begin{equation}
\ket{\psi_{\text{out}}} = \frac{1}{2\sqrt{p_\pm}}\left(\ket{\psi_{\text{in}}}
 \pm U \ket{\psi_{\text{in}}}\right),
\end{equation}
which are the normalized eigenstates of $U$. 

\subsection{Two qubits}
\begin{figure}[ht!]
\hspace{3.5cm}\scalebox{1.4}{
\Qcircuit @C=1em @R=0.8em {
	\lstick{\ket{0}}  &  \gate{H} & \ctrl{1} & \ctrl{2} &\gate{H} & \qw &  \meter \\
	\lstick{}  &  \qw &  \gate{U_1} & \qw & \qw & \qw & \qw\\
	\lstick{} &  \qw & \qw & \gate{U_2} & \qw &  \qw &\qw
	\inputgroupv{2}{3}{0.4em}{1.2em}{\ket{\Phi_{\text{in}}}}\\
}}
\caption{Measuring an operator $U_1 U_2$.} \label{cir:MU2} 
\end{figure}
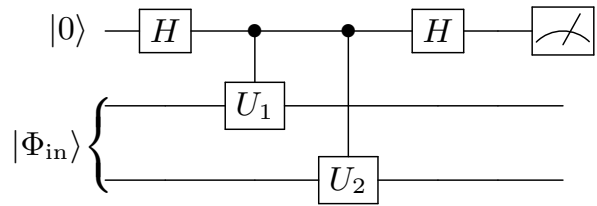

Consider the quantum circuit shown in Fig. \ref{cir:MU2}. 
It is the previous system extended with one qubit and where the observable $U$
is the product of two one-qubit operators $U=U_1 U_2$. 

Here we will show that the measurement of the ancilla again yields the
statistics for a direct measurement of $U$ and that the
final two-qubit state is indeed a corresponding eigenstate. 
In this two-qubit case the eigenvalues $\pm$ are two-fold degenerate.
For each eigenvalue we therefore have a two-dimensional eigenspace. In these
subspaces we choose an orthonormal basis. 

Let us explicitly demonstrate this for the operator $Z_1Z_2$. The results can actually
be used to construct the basis for the general case. The standard basis for the 
two-qubit system is given by
\begin{equation}
\ket{e_1} = \ket{00},  \ket{e_2} = \ket{01},\ket{e_3} = \ket{10},\ket{e_4} = \ket{11}.
\end{equation}
The basis vectors are eigenvectors of $Z_1Z_2$
\begin{align}
Z_1 Z_2 \ket{e_k} &= \ket{e_k} \quad \text{for} \quad k=1,4, \nonumber \\
Z_1 Z_2 \ket{e_k} &= -\ket{e_k} \quad \text{for} \quad k=2,3.
\end{align}
The subspace corresponding to eigenvalue $+1$ is spanned by $\ket{e_1}$ and $\ket{e_4}$
where $\ket{e_2}, \ket{e_3}$ is a basis of the $(-1)$--subspace. In this case, no 
further Gram-Schmidt orthonormalization procedure is necessary. The latter can also
be avoided for $U_1 U_2$ by using these results and the unitary transformations
$V_1, V_2$ defined by
\begin{equation}
 Z_1=V_1 U_1 V_1^\dagger, \quad  Z_2=V_2 U_2 V_2^\dagger,
\end{equation}
cf. (\ref{eq:dia}). Combining the last two equations yields 
\begin{align}
U_1 U_2 V_1^\dagger V_2^\dagger\ket{e_k} &= V_1^\dagger V_2^\dagger \ket{e_k} \quad \text{for} \quad k=1,4, \nonumber \\
U_1 U_2 V_1^\dagger V_2^\dagger\ket{e_k} &= -V_1^\dagger V_2^\dagger\ket{e_k} \quad \text{for} \quad k=2,3,
\end{align}
{\it i.e.}, the orthonormal eigenvectors of $U_1U_2$:
\begin{align}
\ket{u_+^1} &= V_1^\dagger V_2^\dagger \ket{e_1}, \quad
\ket{u_+^2} = V_1^\dagger V_2^\dagger \ket{e_4},  \nonumber \\
\ket{u_-^1} &= V_1^\dagger V_2^\dagger \ket{e_2}, \quad
\ket{u_-^2} = V_1^\dagger V_2^\dagger \ket{e_3}.
\end{align}
Consequently, we get as spectral decomposition of the operator $U=U_1 U_2$,
\begin{equation}
U= \ket{u_+^1}\bra{u_+^1}+\ket{u_+^2}\bra{u_+^2}-
\ket{u_-^1}\bra{u_-^1}-\ket{u_-^2}\bra{u_-^2}.
\label{eq:spec}
\end{equation}
For this degenerate case, the measurement probabilities are given by
\begin{align}
p_+ =&  \left|\braket{u^1_+}{\Phi_{\text{in}}}\right|^2 + \left|\braket{u^2_+}{\Phi_{\text{in}}}\right|^2, \nonumber \\
p_- =&  \left|\braket{u^1_-}{\Phi_{\text{in}}}\right|^2  + \left|\braket{u^2_-}{\Phi_{\text{in}}}\right|^2.
\end{align} 

The circuit shown in Fig. \ref{cir:MU2} without the final measurement
transforms the initial three-qubit state
$\ket 0 \otimes \ket{\Phi_{\text{in}}}$, to
\begin{eqnarray}
 \ket{\Phi_{\text{f}}} &=& \tfrac{1}{2}\ket 0 \otimes (\ket{\Phi_{\text{in}}}+U_1U_2\ket{\Phi_{\text{in}}}) \nonumber \\
 &+& \tfrac{1}{2}\ket 1 \otimes (\ket{\Phi_{\text{in}}}-U_1U_2\ket{\Phi_{\text{in}}}).
 \label{eq:phif}
\end{eqnarray}
The ancilla measurement probabilities easily follow as
\begin{align}
p_0 &= \tfrac{1}{2}\left(1+\bra{\Phi_{\text{in}}} U_1 U_2 \ket{\Phi_{\text{in}}}\right), \nonumber \\
p_1 &= \tfrac{1}{2}\left(1-\bra{\Phi_{\text{in}}} U_1 U_2 \ket{\Phi_{\text{in}}}\right).
\end{align} 
Inserting the spectral decomposition (\ref{eq:spec}) readily yields the desired
result $p_0=p_+, p_1=p_-$. The final two-qubit states can be read off
from (\ref{eq:phif})
\begin{equation}
\ket{\Phi_{\text{out}}} = \frac{1}{2\sqrt{p_\pm}}\left(\ket{\Phi_{\text{in}}}
\pm U_1U_2 \ket{\Phi_{\text{in}}}\right).
\end{equation}
 It is easily checked that these are normalized eigenstates of $U_1U_2$.

If we apply the method to the Bell experiment we need controlled
$Q, R, S, T$ operations. In Sec. \ref{sec:CQRST} it is explicitly derived
how to implement these controlled gates.

\section{Bell experiments III: avoiding the freedom-of-choice loophole}\label{sec:Bell-III}
In order to overcome the freedom-of-choice loophole, a random choice of the
observable from the introduced set $\{QS,RS,RT,QT\}$  in each run is necessary, cf. \cite{Weihs,Scheidl,Rauch}.
Actually motivated by avoiding the locality loophole,
a fast random-number generator is used by \cite{Hensen}.
Here we will demonstrate that random selection is quantum mechanically possible by adding two ancillas, {\it i.e.}, by exploiting
a four-qubit system. Two implementations are foreseen. In the first one, all qubits are finally simultaneously
measured, thereby explicitly confirming that each observable has been measured in
(approximately) 25\% of the runs.
Using the ancilla measurement results, one then can extract the necessary expectation values.

Alternatively, the ancillas are measured first and the choice of the remaining data qubit
gates is determined by these results.
In other words, operations on the data qubits controlled by classical bits are performed.
These two qubits are eventually measured as well.

\subsection{Random choice of the observable}
\begin{figure}[ht!]
\centering
	\scalebox{1.25}{
		\begin{tabular}{c}
\Qcircuit @C=1em @R=.8em
{\lstick{\ket{0}} & \gate{H} & \qw      & \qw      & \qw                & \ctrl{1} & \meter  \\
\lstick{\ket{0}}  & \gate{X} & \gate{H} & \ctrl{1} & \qw                & \gate{H}  & \meter \\
\lstick{\ket{0}}  & \gate{X} & \qw      & \targ    & \gate{R_y(\alpha)} & \gate{R_y(\varphi)}  & \meter\\ 
\lstick{\ket{0}}  & \gate{H} & \qw      & \qw      & \qw                & \ctrl{-1}  & \meter }
		\end{tabular}
}
\caption{Circuit with randomization, the upper and lower qubits are the ancilla qubits; $\varphi=\vartheta-\alpha$, where $\varphi=-\frac{3\pi}{2} \, \text{and} \; \alpha=\frac{\pi}{4}$.} \label{cir:cond} 
\end{figure}
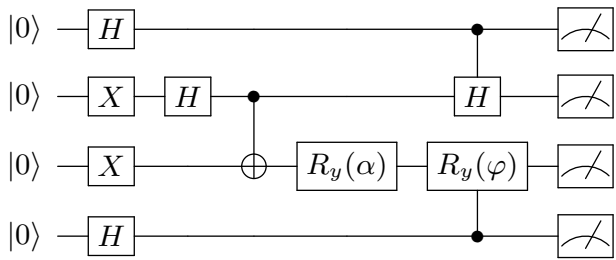

A random choice, with equal probabilities $\tfrac{1}{4}$ can be achieved as follows.
Two ancillas are added to the system and after initialization they are subjected to a Hadamard operation.
The two other qubits are again prepared in a Bell state. Next, the first qubit is {\em conditionally} on the first
ancilla transformed with a Hadamard gate. The second qubit is first rotated by $R_y(\alpha)$. Conditionally on
the second ancilla, another rotation $R_y(\vartheta-\alpha)$ -cancelling $R_y(\alpha)$ and implementing
$R_y(\vartheta)$- is performed.
The corresponding circuit is depicted in Fig. \ref{cir:cond}.

In this way, the desired randomization is achieved within the NISQ device. After the controlled operations
{\it all} qubits are measured. The ancilla results indicate which two-qubit observable has been measured
for the Bell state.  This can verified by analytically computing the final state just before the measurement.
It is given by
\begin{align}
\ket{\Psi_{\text{fin}}} &= (\tf)^3 [ \\&\ket{0} \otimes 
\left( \ket{0} \otimes R_y(\alpha) \ket{1} - \ket{1} \otimes R_y(\alpha) \ket{0}\right) \otimes \ket{0} \nonumber  \\
 +&\ket{0} \otimes 
\left( \ket{0} \otimes R_y(\vartheta) \ket{1} - \ket{1} \otimes R_y(\vartheta) \ket{0}\right) \otimes \ket{1}  \nonumber \\
 +&\ket{1} \otimes 
\left( H \ket{0} \otimes R_y(\alpha) \ket{1} - H \ket{1} \otimes R_y(\alpha) \ket{0}\right) \otimes \ket{0}  \nonumber \\
 +&\ket{1} \otimes 
\left( H\ket{0} \otimes R_y(\vartheta) \ket{1} - H \ket{1} \otimes R_y(\vartheta) \ket{0}\right) \otimes \ket{1} ]. \nonumber
\label{eq:cond}
\end{align} 
Measurements probabilities can of course be calculated from this expression as well. One obtains 
for the results containing the measurement of the $QT$-observable
\begin{align}
p[0000] &= p[0110]= \tfrac{1}{8} \sin^2{(\alpha/2)} \nonumber \\ &= \tfrac{1}{8}\tfrac{1}{2}(1-\cos{\alpha})= \tfrac{1}{4}\tfrac{1}{8}(2-\sqrt{2}), \nonumber \\
p[0010] &= p[0100]= \tfrac{1}{8} \cos^2{(\alpha/2)} \nonumber \\ &=\tfrac{1}{8}\tfrac{1}{2}(1+\cos{\alpha})= \tfrac{1}{4}\tfrac{1}{8}(2+\sqrt{2}).
\end{align} 
First we note that these probabilities indeed add up to 0.25, confirming that in 25\% of the runs this observable is measured.
Furthermore, we obtain the conditional probabilities, {\it i.e.}, only corresponding to the data-qubits
given the ancilla outcome
\begin{align}
p[00] &= p[11]=  \tfrac{1}{8}(2-\sqrt{2}), \nonumber \\
p[01] &= p[10]=  \tfrac{1}{8}(2+\sqrt{2}),
\end{align} 
which add up to one. It yields the expectation value
\begin{eqnarray}
\langle QT \rangle &=& 
 \tfrac{1}{8}(2-\sqrt{2}) \times (1) +
 \tfrac{1}{8}(2-\sqrt{2}) \times (1) \\ && 
 +\tfrac{1}{8}(2+\sqrt{2}) \times (-1) +
 \tfrac{1}{8}(2+\sqrt{2}) \times (-1) \nonumber \\ &=& -\tf, \nonumber
\end{eqnarray}
confirming a result given in (\ref{eq:evalues}). The other expectation values can be obtained 
analogously. For $QS$ we need
\begin{align}
p[0001] &= p[0111]= \tfrac{1}{8} \sin^2{(\vartheta/2)} \nonumber \\&
=\tfrac{1}{8}\tfrac{1}{2}(1-\cos{\vartheta})= \tfrac{1}{4}\tfrac{1}{8}(2+\sqrt{2}), \nonumber \\
p[0011] &= p[0101]= \tfrac{1}{8} \cos^2{(\vartheta/2)} \nonumber \\ &
=\tfrac{1}{8}\tfrac{1}{2}(1+\cos{\vartheta})= \tfrac{1}{4}\tfrac{1}{8}(2-\sqrt{2}),
\end{align} 
which, once again via conditional probabilities, yields
\begin{eqnarray}
\langle QS \rangle &=& 
 \tfrac{1}{8}(2+\sqrt{2}) \times (1) +
 \tfrac{1}{8}(2+\sqrt{2}) \times (1) \\ &&
 +\tfrac{1}{8}(2-\sqrt{2}) \times (-1) +
 \tfrac{1}{8}(2-\sqrt{2}) \times (-1) \nonumber \\ &=& \tf. \nonumber
\end{eqnarray}
For $RT$ and $RS$ we respectively need
\begin{align}
p[1000] &= p[1110]=
\tfrac{1}{4}\tfrac{1}{4}(1+\sin{\alpha})= \tfrac{1}{4}\tfrac{1}{8}(2+\sqrt{2}), \nonumber \\
p[1011] &= p[0101]=
\tfrac{1}{4}\tfrac{1}{4}(1-\sin{\alpha})= \tfrac{1}{4}\tfrac{1}{8}(2-\sqrt{2}),
\end{align} 
and
\begin{align}
p[1001] &= p[1111]=
\tfrac{1}{8}\tfrac{1}{2}(1+\sin{\vartheta})= \tfrac{1}{4}\tfrac{1}{8}(2+\sqrt{2}), \nonumber \\
p[1011] &= p[1101]= 
\tfrac{1}{8}\tfrac{1}{2}(1-\sin{\vartheta})= \tfrac{1}{4}\tfrac{1}{8}(2-\sqrt{2}).
\end{align} 
Hence we obtain 
\begin{equation}
\langle RT \rangle = 
\langle RS \rangle = 
 \tf,
\end{equation}
and confirm 
(\ref{eq:evalues})
and
(\ref{eq:evalues1}).

\subsection{Random choice of the observable using classical bits}
\begin{figure}[ht!]
\centering
	\scalebox{1.25}{
		\begin{tabular}{c}
\Qcircuit @C=1em @R=.8em
{\lstick{\ket{0}} & \gate{H} & \qw      & \qw      & \meter             & \control \cw  \\
\lstick{\ket{0}}  & \gate{X} & \gate{H} & \ctrl{1} & \qw                & \gate{H}  \cwx & \meter \\
\lstick{\ket{0}}  & \gate{X} & \qw      & \targ    & \gate{R_y(\alpha)}  & \gate{R_y(\varphi)} \cwx[1]  & \meter\\ 
\lstick{\ket{0}}  & \gate{H} & \qw      & \qw      & \meter              & \control{-1} \cw  }
		\end{tabular}
}
\caption{Circuit with randomization and `classical' control, the upper and lower qubits are the ancilla qubits; $\varphi=\vartheta-\alpha$.} \label{cir:condm} 
\end{figure}
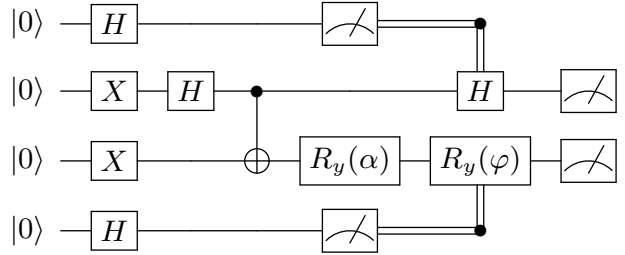

Alternatively, the ancillas are measured before the controlled operations. The latter are then controlled by
the obtained classical bits resulting from the ancilla measurement. Eventually the data qubits are
measured as well.
The corresponding circuit is shown in Fig. \ref{cir:condm}.

It is readily verified that the same statistics and equal expectation values are 
theoretically obtained in this alternative version of the Bell experiment. It reflects
that `measurement commutes with controls', cf. \cite{NC}. These alternatives may be implemented both and the results compared.
Of course, one of the two versions may be preferable in practice.

\subsection{Implementation of the controlled Hadamard and the controlled rotation}
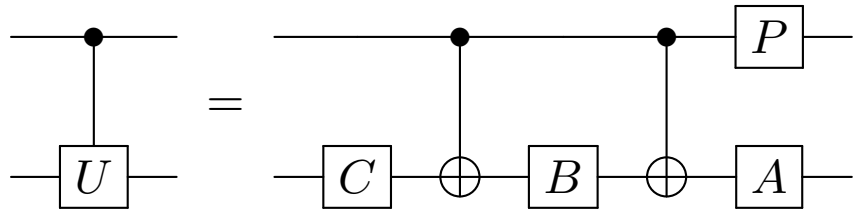
\begin{figure*}[ht!]
\centering
	\scalebox{2}{
		\begin{tabular}{c}
\Qcircuit @C=1em @R=.8em
  { & \ctrl{2} & \qw & & & \qw      & \ctrl{2} & \qw      & \ctrl{2} &\gate{P} &\qw  \\
& & & =& &  & & & & &  \\
   &\gate{U} &\qw & & & \gate{C} & \targ    & \gate{B} & \targ & \gate{A} &\qw}  
		\end{tabular}
}
\caption{Controlled $U$ operation.} \label{cir:CU} 
\end{figure*}

Arbitrary controlled single qubit operations $U$ can be implemented with CNOTs
and single qubit gates \cite{NC}. To this end, one first decomposes $U$ as 
\begin{equation}
U = e^{i\eta} R_z(\beta) R_y(\gamma) R_z(\delta).
\label{eq:decom}
\end{equation}
Next, $U$ is rewritten as
\begin{equation}
U = e^{i\eta} A X B X C \quad \text{with} \quad A B C = \mathcal I,
\end{equation}
where the operators $A, B, C$ are explicitly given by \cite{NC},
\begin{eqnarray}
A &=& R_z(\beta)R_y(\gamma/2), 
 B =R_y(-\gamma/2)R_z(-(\delta+\beta)/2),\nonumber \\
 C &=& R_z((\delta-\beta)/2).
\end{eqnarray}
The circuit implementing the controlled $U$ is shown in Fig. \ref{cir:CU},
where we use the phase operator
\begin{equation}
P=
\begin{pmatrix}
1 & 0 \\
0 & e^{i\eta} 
\end{pmatrix}.
\end{equation}

For the decomposition of the Hadamard gate one has to use 
$\eta=\pi/2, \beta=0, \gamma=\pi/2$ and $\delta=\pi$ in (\ref{eq:decom}).
This can readily be verified by explicit matrix multiplication.
The controlled Hadamard circuit is depicted in Fig. \ref{cir:CUH}.

\begin{figure*}[ht!]
\centering
	\scalebox{1.6}{
		\begin{tabular}{c}
\Qcircuit @C=1em @R=.8em
  { & \ctrl{2} & \qw & & & \qw      & \ctrl{2} & \qw & \qw      & \ctrl{2} &\gate{P} & \qw  \\
& & & =& &  & & & & & \\
   &\gate{H} &\qw & & & \gate{R_z(\tfrac{\pi}{2})} & \targ    & \gate{R_z(-\tfrac{\pi}{2})} & \gate{R_y(-\tfrac{\pi}{4})} &
\targ & \gate{R_y(\tfrac{\pi}{4})} &\qw} 
		\end{tabular}
}
\caption{Controlled $H$ operation.} \label{cir:CUH} 
\end{figure*}
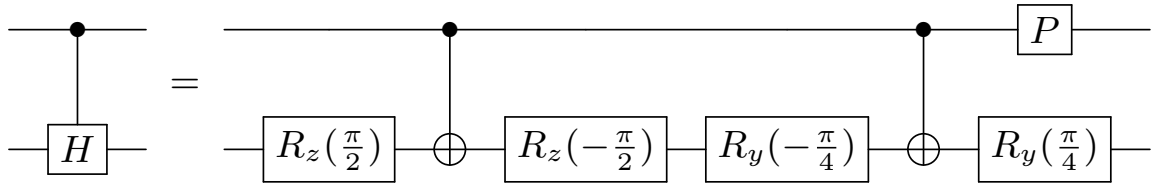


\begin{figure*}[ht!]
\centering
	\scalebox{2.0}{
		\begin{tabular}{c}
\Qcircuit @C=1em @R=.8em
  { & \ctrl{2} & \qw & & & \qw      & \ctrl{2} & \qw   & \ctrl{2}  \qw & \qw &\qw \\
& & & =& &  & & &  & \\
   &\gate{R_y(\varphi)} &\qw & & & \qw & \targ    & \gate{R_y(-\tfrac{\varphi}{2})} &
\targ & \gate{R_y(\tfrac{\varphi}{2})} &\qw} 
		\end{tabular}
}
\caption{Controlled $R_y(\varphi)$ operation.} \label{cir:CURY} 
\end{figure*}
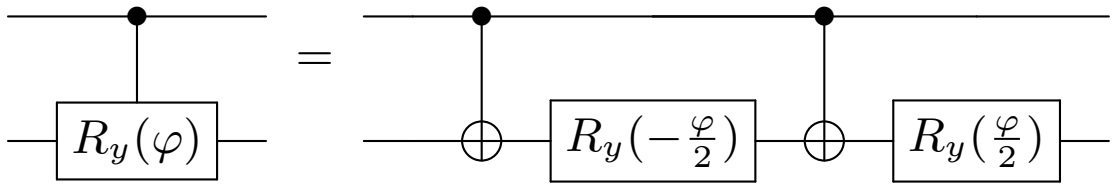

The decomposition (\ref{eq:decom})  of the rotation $R_y(\varphi)$ is trivial: $\eta=\beta=\delta=0$ and 
$\gamma=\varphi$. Consequently, we obtain $A=R_y(\varphi/2), B=R_y(-\varphi/2)$ and the circuit
shown in Fig. \ref{cir:CURY}.

\subsection{Controlled $Q, R, S, T$ operations}\label{sec:CQRST}
If we want to measure the Bell operators by means of an additional ancilla,
we have to implement their controlled version. It is again done  with the decomposition
(\ref{eq:decom}). The following results can readily be verified by explicitly
inserting the angles $\eta,\beta,\gamma, \delta$ and matrix multiplication.
For the observable $Q$ we obtain $\eta=\pi/2, \beta=\pi, \gamma=\delta=0$.
The operator $S$ can be parametrized with
 $\eta=\pi/2, \beta=0., \gamma= \pi/2, \delta=-\pi$.
The observable $T$ follows with 
 $\eta=\pi/2, \beta=\pi, \gamma= \pi/2, \delta=0$.
Finally, we get for the operator $R$ the angles
 $\eta=-\pi/2, \beta=\pi, \gamma= \pi, \delta=0$.
In this way, the controlled operations are feasible on the
quantum simulators and on the NISQ computers.

\section{Bell experiments IV: randomized version and additional ancilla}\label{sec:Bell-IV}
The experiment in the previous section concludes with the measurement of 
four qubits. The $2^4=16$ possible outcomes in principle yield too much
information since only 8 possibilities are of concern. Indeed, for the Bell test,
it is necessary to infer which out of four operators has been measured as well as the corresponding measurement result. Above the results for both, say,
$R$ and $T$ are determined whereas
only the outcome for the product $RT$ is needed.  

Adding an ancilla, which is -just as the other two ancillas- eventually measured, makes the measurement of
 the data-qubits superfluous. In this way, only the products of single
qubit operators are measured and the number of measurements is indeed
reduced to three. It is achieved by means of the technique presented
in Sec. \ref{sec:AI}.

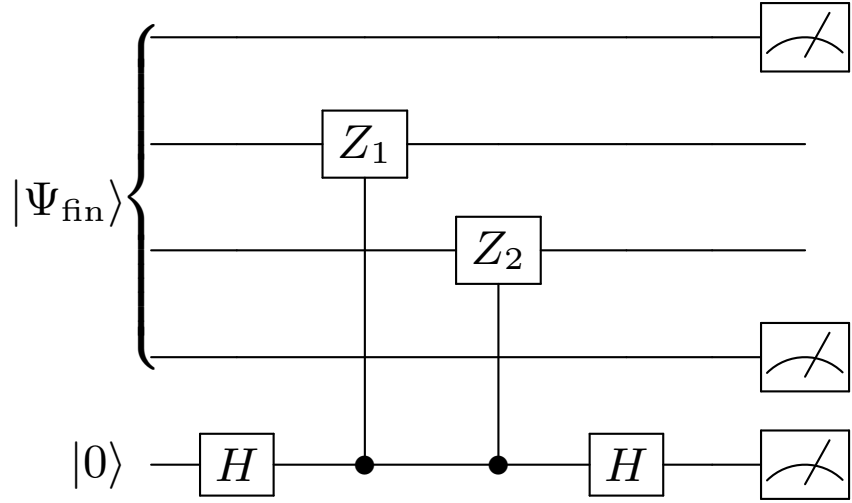
\begin{figure*}[ht!]
	\hspace{3.5cm}\scalebox{2}{
		\Qcircuit @C=1em @R=0.8em {
			\lstick{}  &  \qw &  \qw & \qw & \qw & \qw & \meter \\
			\lstick{}  &  \qw &  \gate{Z_1} & \qw & \qw & \qw &\qw\\
			\lstick{} &  \qw & \qw & \gate{Z_2} & \qw &  \qw &\qw \\
			\lstick{}  &  \qw &  \qw & \qw & \qw & \qw  & \meter\\
			\lstick{\ket{0}}  &  \gate{H} & \ctrl{-3} &\ctrl{-2}&\gate{H} & \qw &  \meter 
			\inputgroupv{1}{4}{0.4em}{3.4em}{\ket{\Psi_{\text{fin}}}}\\
		}}
		\caption{Randomized circuit, additional ancilla.} \label{cir:MU3} 
\end{figure*}

First we employ the circuit depicted in Fig. \ref{cir:cond}, however omitting
the measurements. Its final four-qubit state is given in (\ref{eq:cond}). The circuit is extended with an ancilla and is shown in Fig. \ref{cir:MU3};
note that eventually {\em only} the ancillas are measured.
The latter part of the extended circuit implements -prior to ancilla measurements-
the transformation
\begin{eqnarray}
\ket{\Psi_{\text{fin}}} \otimes \ket 0 \longrightarrow \ket{\Phi_{\text{fin}}} &=&
\tfrac{1}{2}\left(\ket{\Psi_{\text{fin}}}+Z_1Z_2\ket{\Psi_{\text{fin}}}\right)\otimes\ket 0  \nonumber \\
&+&\tfrac{1}{2}\left(\ket{\Psi_{\text{fin}}}-Z_1Z_2\ket{\Psi_{\text{fin}}}\right)\otimes\ket 1. \nonumber \\
&&
\end{eqnarray}
It is convenient to rewrite $\ket{\Psi_{\text{fin}}}$ as
\begin{eqnarray}
\ket{\Psi_{\text{fin}}} &=&\tfrac{1}{2}\left(\ket 0 \otimes \ket{\phi_{00}} \otimes \ket 0
+ \ket 0 \otimes \ket{\phi_{01}} \otimes \ket 1\right) \\
&+& \tfrac{1}{2}\left(\ket 1 \otimes \ket{\phi_{10}} \otimes \ket 0
+ \ket 1 \otimes \ket{\phi_{11}} \otimes \ket 1
 \right) \nonumber,
\end{eqnarray}
where the definitions of the $\ket{\phi_{kl}}$ follow by comparison with (\ref{eq:cond});
note the normalization $\braket{\phi_{kl}}=1$.

Herewith we obtain for the final state of the extended circuit, $\ket{\Phi_{\text{fin}}}$. This final state is given by the following equation:

\begin{widetext}
\begin{align}
\ket{\Phi_{\text{fin}}} =\tfrac{1}{4}[&
\ket 0 \otimes \left(\ket{\phi_{00}} +Z_1Z_2\ket{\phi_{00}}\right)\otimes \ket 0 \otimes \ket 0
+\ket 0 \otimes \left(\ket{\phi_{00}} -Z_1Z_2\ket{\phi_{00}}\right)\otimes \ket 0 \otimes \ket 1 \nonumber \\
 &\ket 0 \otimes \left(\ket{\phi_{01}} +Z_1Z_2\ket{\phi_{01}}\right)\otimes \ket 1 \otimes \ket 0
+\ket 0 \otimes \left(\ket{\phi_{01}} -Z_1Z_2\ket{\phi_{01}}\right)\otimes \ket 1 \otimes \ket 1 \nonumber \\
&\ket 1 \otimes \left(\ket{\phi_{10}} +Z_1Z_2\ket{\phi_{10}}\right)\otimes \ket 0 \otimes \ket 0
+\ket 1 \otimes \left(\ket{\phi_{10}} -Z_1Z_2\ket{\phi_{10}}\right)\otimes \ket 0 \otimes \ket 1 \nonumber \\
&\ket 1 \otimes \left(\ket{\phi_{11}} +Z_1Z_2\ket{\phi_{11}}\right)\otimes \ket 1 \otimes \ket 0
+\ket 1 \otimes \left(\ket{\phi_{11}} -Z_1Z_2\ket{\phi_{11}}\right)\otimes \ket 1 \otimes \ket 1 ].
\end{align}
\end{widetext}
At this point, the $Z$ observables of the ancillas are measured. The results of the
first two ancillas indicates which two-qubit operator is addressed whereas the third outcome yields the corresponding result. We denote the probability that 
the observable $AB$ is (indirectly) measured with result $\pm$ by $p_\pm^{AB}$.
These probabilities can be calculated with the results obtained so far. For the observable $QT$ we get 
\begin{eqnarray}
p_1^{QT}&=&\tfrac{1}{8}\left(\braket{\phi_{00}}+\bra{\phi_{00}}Z_1Z_2\ket{\phi_{00}}\right), \nonumber \\
p_{-1}^{QT}&=&\tfrac{1}{8}\left(\braket{\phi_{00}}-\bra{\phi_{00}}Z_1Z_2\ket{\phi_{00}}\right).
\end{eqnarray}
Hence the probability that this observable is measured indeed equals $\tfrac{1}{4}$. 
The matrix element $\bra{\phi_{00}}Z_1Z_2\ket{\phi_{00}}$ can be calculated by inserting the explicit form of $\ket{\phi_{00}}$ and using the results of Sec. \ref{sec:MQ}.
In this way we find 
\begin{equation}
\bra{\phi_{00}}Z_1Z_2\ket{\phi_{00}} = \tfrac{1}{2}\left(\bra 1 T \ket 1 - \bra 0 T \ket 0
 \right)
 = -\tfrac{1}{2}\sqrt{2},
\end{equation}
and, consequently, 
\begin{equation}
p_1^{QT}=\tfrac{1}{4} \tfrac{1}{4}(2-\sqrt{2}), \quad
p_{-1}^{QT}=\tfrac{1}{4} \tfrac{1}{4}(2+\sqrt{2}). 
\end{equation}
We have separated the selection probability $\tfrac{1}{4}$ and we eventually
confirm the expectation value $ \ev{QT}= -\tfrac{1}{2}\sqrt{2}$.

Analogously we obtain for the observable $QS$ the probabilities
\begin{eqnarray}
p_1^{QS}&=&\tfrac{1}{8}\left(\braket{\phi_{01}}+\bra{\phi_{01}}Z_1Z_2\ket{\phi_{01}}\right), \nonumber \\
p_{-1}^{QS}&=&\tfrac{1}{8}\left(\braket{\phi_{01}}-\bra{\phi_{01}}Z_1Z_2\ket{\phi_{01}}\right),
\end {eqnarray}
confirming the selection probability $\tfrac{1}{4}$. The nontrivial matrix element follows as
\begin{equation}
\bra{\phi_{01}}Z_1Z_2\ket{\phi_{01}} = \tfrac{1}{2}\left(\bra 1 S \ket 1 - \bra 0 S \ket 0
 \right) = \tfrac{1}{2}\sqrt{2},
\end{equation}
which yields
\begin{equation}
p_1^{QS}=\tfrac{1}{4} \tfrac{1}{4}(2+\sqrt{2}), \quad
p_{-1}^{QS}=\tfrac{1}{4} \tfrac{1}{4}(2-\sqrt{2}) ,
\end{equation}
and the expectation value $\ev{QS}= \tfrac{1}{2}\sqrt{2}$.
Next we get for the observable $RT$ the probabilities
\begin{eqnarray}
p_1^{RT}&=&\tfrac{1}{8}\left(\braket{\phi_{10}}+\bra{\phi_{10}}Z_1Z_2\ket{\phi_{10}}\right), \nonumber \\
p_{-1}^{RT}&=&\tfrac{1}{8}\left(\braket{\phi_{10}}-\bra{\phi_{10}}Z_1Z_2\ket{\phi_{10}}\right),
\end {eqnarray}
with matrix element  
\begin{equation}
\bra{\phi_{10}}Z_1Z_2\ket{\phi_{10}} = - \tfrac{1}{2}\left(\bra 1 T \ket 0 + \bra 0 T \ket 1
 \right) = \tfrac{1}{2}\sqrt{2}.
\end{equation}
It gives probabilities
\begin{equation}
p_1^{RT}=\tfrac{1}{4} \tfrac{1}{4}(2+\sqrt{2}), \quad
p_{-1}^{RT}=\tfrac{1}{4} \tfrac{1}{4}(2-\sqrt{2}) ,
\end{equation}
and the resulting expectation value $\ev{RT}= \tfrac{1}{2}\sqrt{2}$.
Finally, we obtain for observable $RS$ the probabilities
\begin{eqnarray}
p_1^{RS}&=&\tfrac{1}{8}\left(\braket{\phi_{11}}+\bra{\phi_{11}}Z_1Z_2\ket{\phi_{10}}\right), \nonumber \\
p_{-1}^{RS}&=&\tfrac{1}{8}\left(\braket{\phi_{11}}-\bra{\phi_{11}}Z_1Z_2\ket{\phi_{10}}\right) ,
\end{eqnarray}
and for the relevant matrix element
\begin{equation}
\bra{\phi_{11}}Z_1Z_2\ket{\phi_{11}} = - \tfrac{1}{2}\left(\bra 1 S \ket 0 + \bra 0 S \ket 1
 \right) = \tfrac{1}{2}\sqrt{2},
\end{equation}
yielding the probabilities 
\begin{equation}
p_1^{RS}=\tfrac{1}{4} \tfrac{1}{4}(2+\sqrt{2}), \quad
p_{-1}^{RS}=\tfrac{1}{4} \tfrac{1}{4}(2-\sqrt{2}) ,
\end{equation}
and the expectation value $\ev{RS}= \tfrac{1}{2}\sqrt{2}$.
Hence we indeed confirm the random selection probabilities and the expectation
values for the four observables in the Bell experiment.

\section{Simulations}\label{sec:sim2}
\subsection{Bell experiments I - QX}\label{sec:sim}
The in Sec. \ref{sec:Bell-I} proposed implementation of the experiment can be tested with the QX simulator of Quantum Inspire \cite{QI}.
To this end, we have programmed the four circuits presented in Sec. \ref{sec:Bell-I}.
The quantum programs have been performed $4 \times 1024$ times, because 1024 is the maximum
number of shots. The following expectation values are obtained with 0.01 standard deviation
\begin{eqnarray}
< QS >  &=& 0.72,\quad <RS>  = 0.70, \nonumber \\
   <RT> &=& 0.72,\quad  <QT> = - .70 .
\end{eqnarray} 
Considering the standard deviation, the Bell inequality (\ref{eq:Bellin}) is clearly violated and our approach confirmed.

To render such simulations more realistic, noise can be included in the QX simulator. 
Only one error model is available at present. This is the ``symmetric depolarizing channel" where the 
per-operation error has to be set. Typical values vary between 0.001 and 0.01; we have chosen 0.005. 
We have run the algorithms including noise $4 \times 1024$ times with the following results with 0.01 standard deviation
\begin{eqnarray}
< QS >  &=& 0.64, \quad <RS>  = 0.67, \nonumber \\  <RT> &=& 0.66,  \quad  <QT> = - .68 .
\label{eq:resqx}
\end{eqnarray} 
All expectation values decrease
(in absolute value) somewhat, which is to be expected since the theoretical values (and
consequently the noiseless results)
correspond to the maximal violation of the Bell inequality. Considering the standard deviation, the
result (\ref{eq:resqx}) still clearly violates the Bell inequality.

\subsection{Bell experiments II - IBM}
The experiment proposed in Sec. \ref{sec:AI}, which only measures one ancilla qubit, has been tested with the IBM simulator \cite{IBM}.
We have programmed the circuit presented in Fig. \ref{cir:MU2} for $U_1 U_2= QS, QT, RS, RT$.
The quantum programs have been performed $4 \times 8192$ times for each observable, with 8192 equal to the maximum
number of shots. The following expectation values are obtained with 0.004 standard deviation
\begin{eqnarray}
< QS >  &=& 0.700, \quad <RS>  = 0.703, \nonumber \\  <RT> &=& 0.709,  \quad  <QT> = -0.712 .
\end{eqnarray} 
Considering the standard deviation, the Bell inequality (\ref{eq:Bellin}) is clearly violated and our approach confirmed. Also, in case a depolarizing error model is included with an error value of 0.005, the Bell inequality is still violated.
	
\subsection{Bell experiments III - QX}
The proposed implementations of the experiments including
random choice of the observables has also been tested with
the QX simulator of Quantum Inspire \cite{QI}.
It is fairly straightforward to program the circuits
shown in Figs. \ref{cir:cond} and \ref{cir:condm}.

The results of both sets simulations confirm the theoretical predictions. For
the circuit in Fig. \ref{cir:cond} which terminates with a joint measurement, one optionally runs
the code without the eventual measurement. In that case
exact, deterministic probability amplitudes are returned by the simulator. Including the measurement
generates quantum noise and running  the algorithm a number of times mimics 
`ideal' experiments. Since the circuit in Fig. \ref{cir:condm} relies on intermediate measurements
of the ancilla a completely deterministic output cannot be generated.

Recall that it is possible to include a depolarizing channel as error model.
More realistic experiments are simulated in this way.
Both algorithms still perform well -in the sense of indeed violating the Bell inequality-
for reasonably low settings of the error parameter, e.g. 0.005.

\subsection{Bell experiments IV - IBM}
The experiment proposed in Sec. \ref{sec:Bell-IV}, which only measures three ancilla qubits, has been tested with the IBM simulator \cite{IBM}.
We have programmed the circuit presented in Fig. \ref{cir:MU3}.
The quantum programs have been performed $16 \times 8192$ times, with 8192 equal to the maximum
number of shots. The following expectation values are obtained with 0.004 standard deviation
\begin{eqnarray}
< QS >  &=& 0.701, \quad <RS>  = 0.712, \nonumber \\  <RT> &=& 0.711,  \quad  <QT> = -0.710 .
\end{eqnarray}
Considering the standard deviation, the Bell inequality (\ref{eq:Bellin}) is clearly violated and our approach confirmed. Also, in case a depolarizing error model is included with an error value of 0.005, the Bell inequality is still violated.

\section{Experiments on the IBM Q Experience}

\subsection{Implementation for IBM backends}

The Bell experiment quantum circuits discussed in Secs. \ref{sec:Bell-I} -- \ref{sec:Bell-IV} have been implemented and have been run first with the simulator backend to verify the implementation and to obtain the right statistics. 

\begin{figure}[h]
	\centering
	\includegraphics[scale=0.5]{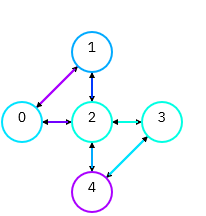}
	\caption{Scheme of available CNOT operations for specific pairs of qubits on the IBM Qx2 backend. The qubits are indicated by the circles; the CNOT operations are indicated by arrows between pairs of qubits (colours of the circles and arrows indicate the gate error rates as given by \cite{IBM}).}
	\label{fig:Qx2}	
\end{figure}

\begin{figure}[h]
	\centering
	\includegraphics[scale=0.5]{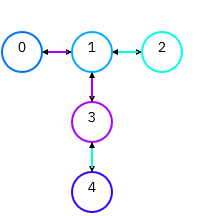}
	\caption{Scheme of available CNOT operations for specific pairs of qubits on the IBM Vigo, Ourense, Essex and Burlington backends. The coloured circles and arrows are explained in the caption of Fig. \ref{fig:Qx2}.}	
	\label{fig:Vigo}
\end{figure}

\begin{figure}[!h]
	\centering
	\includegraphics[scale=0.45]{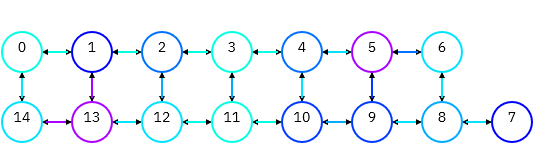}
	\caption{Scheme of available CNOT operations for specific pairs of qubits on the IBM Melbourne backend. The coloured circles and arrows are explained in the caption of Fig. \ref{fig:Qx2}.}	
	\label{fig:Melbourne}
\end{figure}

We have considered one 15-qubit and several 5-qubit IBM hardware backends available for users to perform experiments. An important aspect of these hardware backends is that not all qubits are connected to each other. So two-qubit operations, like CNOT, will be restricted to specific pairs of qubits. This is schematically shown in Figs. \ref{fig:Qx2} -- \ref{fig:Melbourne}.

From the available CNOT operations in these figures it can be seen that the Bell experiment quantum circuits in Sec. \ref{sec:Bell-I} and \ref{sec:Bell-III} can be run on each of these backends. The Bell experiment quantum circuits in Sec. \ref{sec:AI} can only be run on the Qx2 backend. The Bell experiment quantum circuit in Sec. \ref{sec:Bell-IV} cannot be run on any of these backends, it can only be simulated.

\subsection{Results}
The results of the experiments on the IBM Q Experience are given in Tables \ref{tab:IBM-I} -- \ref{tab:IBM-III}. They show the the expectation values for the observables $QS, QT, RS$ and $RT$, as well as the CHSH inequality value $CHSH =$ $ <QS>+<RS>+<RT>-<QT>$. The expectation values in Tables \ref{tab:IBM-I} and \ref{tab:IBM-II} are obtained by using 10x8192 shots for each observable; The expectation values in Table \ref{tab:IBM-III} are obtained by using 20x8192 shots.
The errors for the observables are the standard deviations of the sampled expectation values. E.g. for observable $QS$, the standard deviation is given by
\begin{equation}
\sigma_{<QS>}=\sqrt{\frac{<(QS)^2>-<QS>^2}{N-1}}=\sqrt{\frac{1-<QS>^2}{N-1}}, 
\end{equation}
where $N$ is the number of observations, given by the number of runs multiplied with the number of shots per run. 

Considering the values for $CHSH$ and their standard deviations in Tables \ref{tab:IBM-I} -- \ref{tab:IBM-II}, the Bell inequality (\ref{eq:Bellin}) is clearly violated, except for the Burlington backend. 
For Table \ref{tab:IBM-III}, we used the Bell algorithm from Sec. \ref{sec:Bell-III}, which exploits additional qubits and quantum gates. Here the Bell inequality (\ref{eq:Bellin}) is only violated, for the Qx2 and Vigo backends. 

\begin{table*}
	\caption{Expectation values for observables $QS$, $Q$T, $RS$ and $RT$, according to Sec. \ref{sec:Bell-I}.}	
	\begin{center}
		\begin{tabular}{ l r r r r r}
			\hline \hline
			IBM Backend & $<QS>$ & $<QT>$ &$<RS>$ & $<RT>$ & $CHSH$\\ \hline
			Qx2 & 0.666$\pm$0.003 & -0.511$\pm$0.003 & 0.697$\pm$0.003 & 0.660$\pm$0.003 & 2.533$\pm$0.005\\ 
			Vigo & 0.660$\pm$0.003 & -0.644$\pm$0.003 & 0.638$\pm$0.003 & 0.656$\pm$0.003 & 2.598$\pm$0.005\\ 
			Ourense & 0.659$\pm$0.003 & -0.639$\pm$0.003 & 0.625$\pm$0.003 & 0.654$\pm$0.003 & 2.576$\pm$0.005\\ 
			Essex & 0.633$\pm$0.003 & -0.503$\pm$0.003 & 0.533$\pm$0.003 & 0.632$\pm$0.003 & 2.301$\pm$0.006\\ 
			Burlington & 0.524$\pm$0.003 & -0.347$\pm$0.003 & 0.348$\pm$0.003 & 0.560$\pm$0.003 & 1.779$\pm$0.006\\ 
			Melbourne & 0.735$\pm$0.002 & -0.503$\pm$0.003 & 0.475$\pm$0.003 & 0.711$\pm$0.003 & 2.425$\pm$0.006\\ \hline
			\hline
		\end{tabular}
	\end{center}
	\label{tab:IBM-I}	
\end{table*}

\begin{table*}
	\caption{Expectation values for observables $QS$, $Q$T, $RS$ and $RT$, according to Sec. \ref{sec:AI}}	
	\begin{center}
		\begin{tabular}{ l r r r r r}
			\hline \hline
			IBM Backend & $<QS>$ & $<QT>$ &$<RS>$ & $<RT>$ & $CHSH$\\ \hline 
			Qx2 & 0.568$\pm$0.003 & -0.586$\pm$0.003 & 0.554$\pm$0.003 & 0.589$\pm$0.003 & 2.297$\pm$0.006\\ \hline
			\hline
		\end{tabular}
	\end{center}
	\label{tab:IBM-II}	
\end{table*}

\begin{table*}
	\caption{Expectation values for observables $QS$, $Q$T, $RS$ and $RT$, according to Sec. \ref{sec:Bell-III}}	
	\begin{center}
		\begin{tabular}{ l r r r r r}
			\hline \hline
			IBM Backend & $<QS>$ & $<QT>$ &$<RS>$ & $<RT>$ & $CHSH$\\ \hline 
			Qx2 & 0.599$\pm$0.004 & -0.455$\pm$0.005 & 0.567$\pm$0.004 & 0.572$\pm$0.004 & 2.193$\pm$0.009\\ 
			Vigo & 0.692$\pm$0.004 & -0.470$\pm$0.004 & 0.516$\pm$0.005 & 0.644$\pm$0.004 & 2.342$\pm$0.008\\ 
			Ourense & 0.381$\pm$0.005 & -0.504$\pm$0.005 & 0.525$\pm$0.004 & 0.494$\pm$0.004 & 1.904$\pm$0.009\\ 
			Essex & 0.515$\pm$0.004 & -0.275$\pm$0.005 & 0.405$\pm$0.005 & 0.514$\pm$0.004 & 1.709$\pm$0.009\\ 
			Burlington & 0.338$\pm$0.005 & -0.118$\pm$0.004 & 0.284$\pm$0.005 & 0.438$\pm$0.004 & 1.178$\pm$0.010 \\ 
			Melbourne & 0.339$\pm$0.005 & -0.425$\pm$0.004 & 0.531$\pm$0.004 & 0.408$\pm$0.004 & 1.763$\pm$0.009\\ \hline
			\hline
		\end{tabular}
	\end{center}
	\label{tab:IBM-III}	
\end{table*}


\subsection{Analysis: quantum noise}
The results obtained are somewhat surprising in the sense that the absolute values of the expectation
values are systematically lower than theory predicts. A possible reason for deviations
and the systematics may be quantum noise. In \cite{NC}, it is  described in the framework of quantum operations
and explicitly modeled in terms of various error channels. Such a channel is formulated with a number
of quantum operations $E_k$ which defines the following transformation on an arbitrary density matrix $\rho$:
\begin{equation}
\mathcal E(\rho)  = \sum_k E_k \rho E_k^\dagger.
\end{equation}
In our case, the density matrix corresponding to the pure Bell state (\ref{eq:Bell})
is given by
\begin{equation}
\rho = \frac{1}{2}
\begin{pmatrix}
0 & 0 & 0 & 0  \\
0 & 1 & -1 & 0  \\
0 & -1 & 1 & 0  \\
0 & 0 & 0 & 0  
\end{pmatrix}.
\end{equation}
After having calculated the transformed density matrix for a specific channel, the
modified expectation value for an operator $A$ follows as
\begin{equation}
\expval A = \trace{[A\mathcal E(\rho)]}
\end{equation}
The quantum operations for the following one-qubit channels are explicitly given in \cite{NC}:
{\em bit flip, phase flip, bit-phase flip, amplitude damping} and the {\em generalized amplitude
damping} channel. All depend on a certain probability that the error occurs; the generalized
amplitude damping is parametrized in  terms of two probabilities. 
In the subsequent analysis we assume that these operators act on the first one of the
two data qubits. Furthermore, the {\em depolarizing channel} touching a $d$-dimensional
system is considered in \cite{NC} and also here for the two qubits in the Bell state.
Below the respective channels are denoted by B, P, BP, A, GA and D.

Given the explicit forms of the quantum operations for the various channels, the computations are straightforward. 
Note that the one-qubit operators  need to be extended to the two-qubit space:
\begin{equation}
E^{[1]} \rightarrow E^{[1]} \otimes \mathcal I^{[2]}. 
\end{equation}
Omitting intermediate results we merely present the final expressions in Table \ref{tab:Channels},
in terms of the probabilities $p, p_d$ and $\sin^2{\theta}$, where $1-p$ is the probability of 
a bit, phase or combined bit-phase flip. The probability for complete depolarization is given 
by $p_d/4$ and $\sin^2{\theta}$ is the probability related to amplitude damping, cf. \cite{NC}. In 
the generalized amplitude damping an additional probability appears \cite{NC}. Our results, however,
are independent of it. 
\begin{table*}
	\caption{Effects of quantum noise for the various channels.} 
	\begin{center}
		\begin{tabular}{ l r r r r}
			\hline \hline
	Channel& $\sqrt{2}\expval{QS}$ & $\sqrt{2} \expval{QT}$ &$\sqrt{2}\expval{RS}$ & $\sqrt{2}\expval{RT}$ \\ \hline 
	B & $2p-1$ & $1-2p$ & $1$ & $1$ \\ 
	P & $1$ & $-1$ & $2p-1$ & $2p-1$ \\ 
	BP & $2p-1$ & $1-2p$ & $2p-1$& $2p-1$ \\ 
	A & $\cos^2{\theta}$ & $-\cos^2{\theta}$ & $\cos{\theta}$ & $\cos{\theta}$ \\ 
	GA & $\cos^2{\theta}$ & $-\cos^2{\theta}$ & $\cos{\theta}$ & $\cos{\theta}$ \\ 
	D & $1-p_d$ & $p_d-1$ & $1-p_d$ & $1-p_d$  \\ \hline
			\hline
		\end{tabular}
	\end{center}
	\label{tab:Channels}	
\end{table*}
Indeed we see that quantum noise always, that is in all channels, lowers the absolute values of the expectation.
This is in accordance with our experimental results on the different IBM backends. 
At this point it is however impossible to draw further conclusions on the relative
importance of the various channels.

\section{Conclusion and outlook}
Quantum algorithms to be implemented on few-qubit NISQ computers
are developed and tested. The goal is to demonstrate the violation of a 
Bell inequality \cite{Bell,Yurke,NC}. First, we have used the QX simulator of Quantum
Inspire \cite{QI} and the simulator of IBM Q Experience \cite{IBM} and have confirmed the Bell violation using our proposed quantum circuits in simulations.
The violation is resilient to the inclusion of limited amounts of additional noise by means of the depolarizing channel error model.

Secondly, we have performed actual quantum computations on real NISQ devices, namely
IBM Q Experience. The algorithms have been
implemented and the obtained results are 
confirming the violation of the Bell inequality by various standard deviations depending on the type of algorithm and type of device.  
Concomitantly, these experiments serve as a test for the IBM quantum chips, in particular
their ability to generate entanglement and perform reliable single-shot measurements.
We find that the algorithms that exploit additional qubits and quantum gates produce worse results on these chips. For the Qx2 chip we find violation of the Bell inequality for three types of Bell algorithms.

Of course, the question arises what are the consequences for local-realism. 
On such small NISQ devices locality loopholes obviously cannot be ruled out as in \cite{Hensen}. The detection loophole, however, is avoided as in principle has already
been shown in \cite{Ansmann}. Finally, the freedom-of-choice loophole is evaded
by the random choice of the observable, {\it i.e.}, the measurement basis. It is
implemented quantum mechanically in our algorithms to be executed on a four-qubit system, 
exploiting two qubits as controls. Hence no further random-generator is necessary. 
   
In the nearby future, we aim to run our algorithms also on a spin qubit
NISQ computer to confirm our present results using semiconductor qubits.
 One can also pursue the issue of locality
exploiting several NISQ devices.

\section*{Acknowledgements} \nonumber
This research has been supported by the Quantum Technology research group of TNO.  The authors also thank M.J. Woudstra for a critical reading of the manuscript.

\end{document}